\documentclass[aps,prb,twocolumn,showpacs,achemso,amsmath,amssyb]{revtex4}

\usepackage{graphicx,epsfig}

\begin{document}

\title{InAs/InP single quantum wire formation and emission at 1.5 $\mu$m}

\author{B. Al\'en}
\email[]{benito@imm.cnm.csic.es}

\author{D. Fuster}

\author{Y. Gonz\'alez}

\author{L. Gonz\'alez}

\affiliation{IMM, Instituto de Microelectr\'onica de Madrid (CNM, CSIC), Isaac Newton 8,
 28760 Tres Cantos, Madrid, Spain.}

\author{J. Mart\'\i{}nez-Pastor}

\affiliation{ICMUV, Instituto de Ciencia de Materiales, Universidad de Valencia,
  P.O. Box 2085, 46071 Valencia, Spain.}

\date{\today}

\begin{abstract}

Isolated InAs/InP self-assembled quantum wires have been grown using \textit{in situ}
accumulated stress measurements to adjust the optimal InAs thickness. Atomic force
microscopy imaging shows highly asymmetric nanostructures with average length exceeding
more than ten times their width. High resolution optical investigation of as-grown
samples reveals strong photoluminescence from individual quantum wires at 1.5 $\mu$m.
Additional sharp features are related to monolayer fluctuations of the two dimensional
InAs layer present during the early stages of the quantum wire self-assembling process.

\end{abstract}

\pacs{81.07.Vb, 78.67.Lt, 73.21.Hb}

\maketitle

Due to their smaller lattice and energy gap mismatch, InAs nanostructures grown on InP
substrates are initially much better suited to exploit the long wavelength telecom
windows than their GaAs counterparts. However, like the latter, to fulfill the
requirements of solid-state based quantum information technologies, they have to be
produced with low areal densities and high emission efficiencies. In this direction,
emission from single self-assembled InAs/InP quantum dots (QDs) and dashes has been
reported.~\cite{Mensing2003,Chithrani2004,Takemoto2004,Salem2005,Saint-Girons2006} Under
certain growth conditions and epitaxial methods, self-assembled quantum wires (QWRs) can
also be obtained depositing InAs on
InP.~\cite{Li1999,Walther2000,Gonzalez2000,Gutierrez2001,Schwertberger2003} Depending on
the wire size and composition, emission tuning capability in the 1.2~-~1.9 $\mu$m range
has been demonstrated, and both, continous wave and time-resolved optical
characterization, have been performed on high density QWR
arrays.~\cite{Fuster2004,Fuster2005,Alen2001,Fuster2005b} In such situation, long
wavelength emission from single QWRs would be desirable. However, for the time being,
single quantum wire optical spectroscopy has been restricted mainly to GaAs/AlGaAs
nanostructures emitting on the visible range. The best examples are cleaved edge
overgrown T-shaped QWRs,~\cite{Harris1996,Akiyama2003} V-grooved QWRs grown on nonplanar
substrates,~\cite{Crottini2001,Guillet2003b} and recently reported, vertical QWRs grown
on inverted tetrahedral pyramids etched away on GaAs substrates.~\cite{Zhu2006}

By using \textit{in situ} accumulated stress measurements, the initial phases of the
self-assembling process can be monitored during solid source molecular beam epitaxy (MBE)
of InAs on InP(001).~\cite{Garcia2001} We have exploited this fact to choose the optimal
InAs deposited thickness and tailor the QWR areal density from zero to full coverage of
the sample surface. A detailed study of the initial stages of QWR formation will be
provided elsewhere. In the following, using a low areal density sample, we show
experimental evidences of single InAs/InP QWR emission at $\sim$1.55 $\mu$m.

\begin{figure}[t]
\includegraphics[width=58 mm]{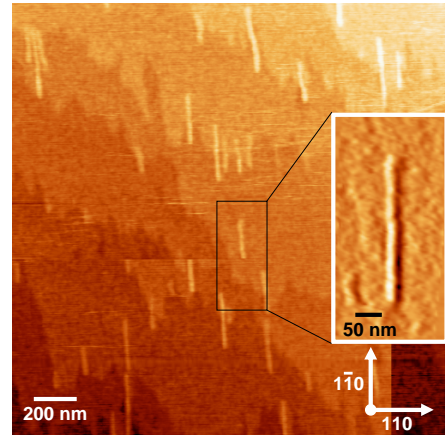}%
\caption{2$\times$2 $\mu$m AFM image of the uncovered sample surface after deposition of
1.5 ML of InAs. Self-assembled quantum wires and one-ML-high InAs steps appear seldom
distributed over the flat InAs surface. Inset: Zoom over a region containing a long QWR
(first derivative image mode).} \label{Fig2}
\end{figure}

The sample studied in this work consist of 1.5 ML of InAs (1 ML InAs~=~3.03
$\textrm{\AA}$) deposited on InP (001) at $r_g$=0.5 ML/s , substrate temperature
$T_S$=480 $^\circ$C and beam equivalent pressure $BEP(As_4)$=6.4$\times$10$^{-6}$ mbar. A
180 nm-thick buffer layer was deposited and before the In cell was opened again for InAs
growth, the InP surface was exposed to As flux during 3 s. After QWRs formation, a
20-nm-thick cap layer was deposited at the same $r_g$ to allow for optical investigation.
In addition, an uncapped sample was grown under the same conditions for atomic force
microscopy (AFM) investigation.

A conventional low magnification setup and non-resonant excitation at 514.5 nm have been
used to study the ensemble averaged photoluminescence (PL) spectrum at 15 K. The
collected light was detected synchronously by a Ge cooled photodetector attached to a
0.22 m focal length monochromator. The optical emission spectra at the single QWR level
have been investigated using a confocal microscope working at 4.2 K (Attocube CFM-I). Low
level light detection has been achieved using a nitrogen cooled InGaAs focal plane array
(Jobin-Yvon IGA-3000) attached to a 0.5 m focal length grating spectrograph. For this
study, light excitation at 950 nm was delivered by a Ti-Saphire tunable laser.

Figure~\ref{Fig2} shows an AFM image of the uncovered sample surface. Elongated
nanostructures appear randomly distributed with an average areal density of $\sim6$
$\mu$m$^{-2}$. Statistics performed in more than 50 of such nanostructures reveal average
widths and lengths of 21$\pm$3 nm and 185$\pm$50 nm, respectively. We note that, although
the average aspect ratio is $\frac{L_y}{L_x}\sim$~8, QWRs with larger anisotropy ($>$~16)
can be easily found as shown in the inset of Figure~\ref{Fig2}. The low QWR areal density
allows a direct analysis of the surface morphology underneath them. We observe a flat
surface with one-monolayer-high steps due to the unintentional miscut angle of the
InP(001) wafer ($\sim$0.06$^\circ$ off). In most cases, nucleation of the self-assembled
nanostructures occurs preferentially in the step edges. However, a number of QWRs have
been also found in flat areas far from any terrace border. The most remarkable
observation is that, despite the orientation of the neighbor terrace step, QWRs growth
occurs always along the [1$\overline{1}$0] direction. To this respect, any dependence
among the QWR formation mechanism and the precise wafer miscut angle can be ruled out in
our case.

\begin{figure}[t]
\includegraphics[width=66.5 mm]{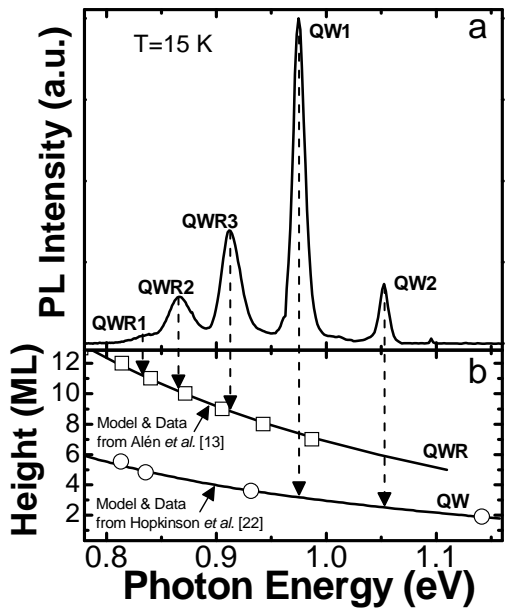}%
\caption{a) Photoluminescence spectrum obtained at 15 K using an standard low
magnification setup. b) PL energy vs. QW/QWR height dispersion curves have been extracted
from previous studies. The energy position of the present emission bands are indicated by
down-arrows in the corresponding curve.} \label{Fig3}
\end{figure}

It should be noted that with 1.5 ML of InAs deposited, plus an additional ML resulting
from the As/P exchange process,~\cite{Gonzalez2004} the sparse QWR nucleation can not
exhaust the initial InAs two-dimensional layer. In these conditions, a thin wetting layer
must remain over the sample surface like usually observed for other self-assembled
nanostructures. These low density QWR samples show optical features that can be related
to an extended wetting layer of fluctuating thickness consistent with the amount of InAs
deposited and the AFM characterization shown above. We must remark that the corresponding
bands (QW1-2) dominate the PL spectrum for low InAs coverage and low QWR densities, but
are effectively quenched when the QWR density is higher. They are hence strongly related
to the incomplete self-assembling process occurring in this case.

The sample PL spectrum consists of five emission bands independently of the excitation
power. A multimodal hypothesis can explain this behavior assuming that each band
corresponds to QWRs or quantum wells (QWs) of a given constant
height.~\cite{Alen2001,Fuster2004} Figure~\ref{Fig3}(b) contains theoretical predictions
and experimental data for both InAs/InP QWRs studied before by the
authors,~\cite{Alen2001} and InAs/InP strained QWs grown and characterized by Hopkinson
\textit{et al}.~\cite{Hopkinson1991} The solid lines represent the expected energy
dispersion as calculated by the respective authors. We start associating the lowest
energy band in the spectrum of Figure~\ref{Fig3}(a) to especially thick QWRs. They are
rare in the ensemble distribution in view of the weak intensity of the corresponding PL
peak (QWR1, 2~$\%$ of the total integrated intensity). Comparing with the QWR dispersion
curve shown in Figure~\ref{Fig3}(b), we observe that its emission energy, marked with a
downarrow, agrees well with 11-ML-thick quantum wires. Peaks labelled as QWR2 (13~$\%$)
and QWR3 (26~$\%$) also agree with data collected in high density samples for 10 ML and 9
ML thick nanostructures, respectively. The series is suddenly cut off since the peak for
8 ML do not appear in our spectrum. Instead, sixty percent of the total integrated
intensity is emitted in peaks labelled QW1 and QW2 in this sample. The peak energies for
bands QW1 and QW2 are in between those measured for 5 and 10 $\textrm{\AA}$ thick
InAs/InP QWs represented in Figure~\ref{Fig3}(b).~\cite{Hopkinson1991} As expected, they
are also narrower, with full widths at half maximum around FWHM$\sim$15 and 12 meV,
respectively, than the inhomogeneously broadened QWR emission bands (FWHM$>22$ meV). As
explained above, we tentatively assign these two peaks to radiative recombination in
monolayer fluctuations of a residual InAs wetting layer.

\begin{figure}[t]
\includegraphics[width=66.5 mm]{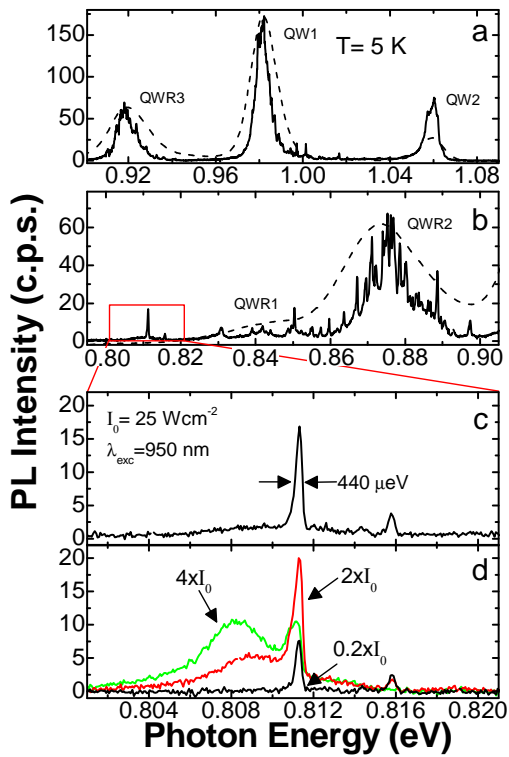}%
\caption{a) and b) High resolution PL spectra measured at 5 K (solid lines) are compared
with the inhomogeneously broadened emission bands (dashed lines). On the low energy tail,
the emission from single QWRs can be studied individually as shown in b) and c). d)
Evolution of the single QWR PL spectrum obtained increasing the excitation power
density.} \label{Fig4}
\end{figure}

In Figure~\ref{Fig4}, the high resolution PL spectrum obtained with our confocal setup at
5 K is compared with the ensemble averaged spectrum. The initial broad bands (dashed
lines) are resolved into different sharp resonances at each specific location in the
sample surface. This occurs both for the low energy components, QWR1-3, as for the higher
energy bands, QW1 and QW2. For the latter, it reveals that local fluctuations of the
wetting layer thickness give rise to interface islands of varying lateral extent.
However, our more relevant findings are in the low energy part of the spectrum. Without
additional sample processing, in this region, the size inhomogeneity and the low areal
density can be exploited to study the emission of specially large single QWRs
[Figure~\ref{Fig4}(b), rectangle].

Laser excitation below the InP bandgap produces intense and sharp emission lines near
1.55~$\mu$m. For the spectrum shown in Figure~\ref{Fig4}(c), the resonance linewidth is
$\sim$440 $\mu$eV, which is around twice our spectral resolution. Indeed, the main
emission peak is strongly asymmetric towards lower energies. Assuming a radiative limited
linewidth alone, which in any case should produce a lorentzian lineshape, it would imply
an unreasonably small value of $\sim$1.5 ps for the exciton radiative lifetime. Our
result rather suggest that the homogeneous linewidth in our QWRs is not radiative limited
as usually reported for QDs. Moreover, increasing the laser intensity, the emission
saturation behavior depicted in Figure~\ref{Fig4}(d) does not follow the typical
exciton-biexciton recombination dynamics commonly observed for InAs/InP
QDs.~\cite{Mensing2003,Chithrani2004,Takemoto2004,Salem2005,Saint-Girons2006} The large
aspect ratio of these QWRs suggest that they could be on a different confinement regime.
Our present results indicate that this is the case concerning their non-linear emission
properties. Two regimes can be identified depending on the saturation behavior of the
main peak. Initially, the integrated intensity of the sharp resonance increases almost
linearly with an exponent $m$=0.98. At the lowest powers, the minimum achieved linewidth
is limited by our spectral resolution. However, raising the excitation density slightly,
the resonance asymmetry becomes patent even before the peak intensity has got completely
saturated. The spectrum in Figure~\ref{Fig4}(c) is representative of this regime. Further
power increase comes along with the appearance of a noticeable broad sideband shifted in
energy from the main peak. In this regime, the sideband integrated intensity raises
sublinear with $m$=0.8 and shifts to lower energies while the main peak gets quenched at
its original energy position. Finally, at higher power (not shown), or under non-resonant
excitation, each sharp feature disappears embedded by its associated broad sideband
(FWHM$\sim$4 meV).

In semiconductor QWRs, the many body correlations present among excitons and free
electron-hole pairs (EHP) lead to strongly modified optical absorption and gain spectra
when increasing the EHP density.~\cite{Sarma2000,Wang2001,Huai2006} The electron and hole
Coulomb interaction is screened by the photogenerated plasma causing the reduction of the
exciton oscillator strength. This reduction partially cancels the band gap
renormalization effect leaving the exciton energy unchanged when increasing power. Our
results shown in Figure~\ref{Fig4}(d) evidence a clear reduction of the exciton resonance
intensity at constant energy in favor of an increasingly broad emission band appearing at
nearby energies. These spectral features must be explained assuming for our
nanostructures a quasi-1D confining potential consistent with their elongated shape. They
are comparable to similar optical features reported for other semiconductor QWRs based in
the GaAs/AlGaAs heteroepitaxial system.~\cite{Akiyama2003,Guillet2003b} Compared with
them, the QWRs presented here can be fabricated by a self-assembled process in nominally
flat InP(001) substrates, avoiding any prepatterning and/or regrowth steps, and making
easier their implementation in current photonic and optoelectronic telecommunication
devices.

To conclude, we have demonstrated that isolated InAs self-assembled quantum wires can be
grown in nominally flat InP(001) substrates using conventional epitaxial techniques.
Emission spectra from individual QWRs have been reported in the wavelength region of
interest for optical communications. Preliminar investigation of their optical properties
reveals spectral signatures of the many-body carrier interactions occurring in 1D
systems.

The authors gratefully acknowledge financial support by the Spanish MEC and CAM through
projects NANOSELF2, NANIC and QUOIT (TEC-2005-05781-C03-01/03, NAN2004-09109-C04-01/03
and S-505/ESP/000200), and by the European Commission through SANDIE Network of
Excellence (No. NMP4-CT-2004-500101).


\end{document}